\documentclass[aps,prc,twocolumn,floatfix]{revtex4}
\usepackage{epsfig}

\begin{document}

\title{Comment on Afshar's expriments}
\author{Daniel Reitzner}
\affiliation{Research Center for Quantum Information, Institute of Physics, Slovak Academy of Sciences, D\'ubravsk\'a cesta 9, 811 45 Bratislava, Slovakia}
\date{\today}

\begin{abstract}
Results of the experiments carried out in \cite{Afshar_old} and \cite{Afshar_new} are reviewed and their interpretation by the authors is questioned. Arguments are supported by numerical simulations.
\end{abstract}

\maketitle

\section{Introduction}

In \cite{Afshar_new} a two-slit experiment supposedly showing the violation of the priciple of complementarity is realized. Authors suggest that a which-way information is obtained from the position of incident photon/light while at the same time the existence of interference is testified. In several articles this experiment is being criticised. These critiques are based usually on explanations dwelling deep inside the basics of quantum mechanics. Here we present numerical results showing that there is no need to go so deep to show that the principle of complementarity hold also in this case. In \cite{Tabish} a simple analytic analysis is shown to support our arguments and numerical simulation.

The claim of Afshar, that the law of conservation of linear momentum compels us to accept that a photon in a particular spot on the photosensitive surface must have originated from the corresponding pinhole, is indeed correct. While having only one slit open the momentum of the photon forces it to end in a state from which an information about pinhole can be obtained. However having both slits open transverse momentum of photon (as a quantum-mechanical object) is zero and at the end must stay zero. The final state of the photon is thus symmetric and holds no evidence about the slits. In other words the two-peaked distribution is an interference pattern and the photon behaves as a wave and exhibits no particle properties until it hits the plate. As a result a which-way information can never be obtained in this way.

\section{Numerical simulation}

Tu support previous arguments we performed numerical simmulations following set-up used in \cite{Afshar_old}. In the simulation process we utilized Huygens-Fresnel principle in the following form. The wave function on a chosen plane (surface) can be expressed in the form
\begin{equation}
\Psi(r)\propto\int\limits_\sigma \Psi_0(\sigma')\frac{e^{ikr}}{r}d\sigma',
\label{integral}
\end{equation}
where $\sigma$ represents the surface of the previous plane from which the wave originates, e.g.~for the wave function on the first interference plane $\sigma$ represents the surface of the two pinholes; $r$ represents the distance from the point of origin and $\Psi_0$ is the initial wave-function (on the surface of the slits being constant). To succesfully simulate the propagation of the wave we alsohave to know, how to implement the lens. Light passing through a lens changes its phase depending on the distance $y$ from the center of the lens, since the thickness of the lens depends on this distance. This phase-shift (up to the unsignificant constant) can be expressed as:
\begin{equation}
\delta(y)=-2k\sqrt{4f^2+y^2},
\end{equation}
where $f$ is the focal length and $k=\frac{2\pi}{\lambda}$ is the circular wave-number of the wave, with $\lambda$ being its wave-length. Including this in the simulation is done by changing the exponent in the intetgral Eq.(\ref{integral}) to $ikr+\delta(y)$ and $y$ is the position on the originating plane --- lens. According to previous scheme we obtained following results.

\section{Results}

\begin{figure}
\includegraphics[scale=0.32,angle=-90]{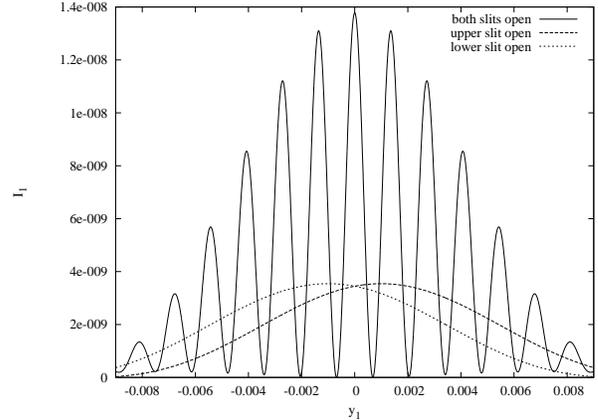}
\caption{\label{fig01}Results for light passing through selected slits (both, upper, or lower one) --- the intensity of light as a function of position on the first interference plate. With both slits open the interference pattern is clearly visible.}
\end{figure}

First we simulated the intensity of light on the first photosensitive surface (which is in the experiment afterwards taken away). If the light is passing only throug one slit (Fig.~\ref{fig01}), we see that the resulting intensity of light --- in quantum mechanical sense the probability density function of the incident photon --- has only one peak. Opening both slits creates an interference pattern that is clearly observable and corresponds also quantitatively to Afshar's results.

\begin{figure}
\includegraphics[scale=0.32,angle=-90]{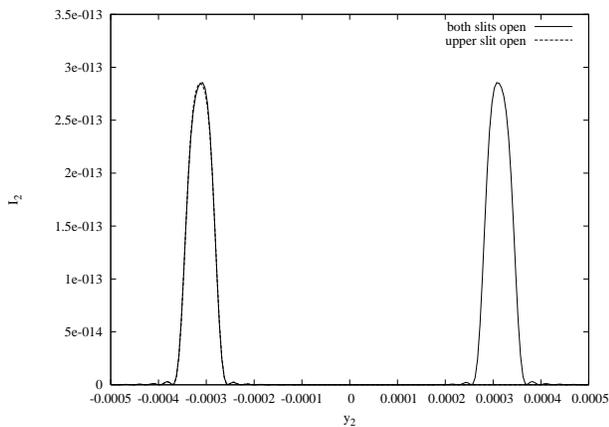}
\caption{\label{fig02}Results for light passing through selected slits (both or only upper one) and lens. As geometrical optics predicts the image of opened slits is obtained in the predicted positions. However which-way information cannot be obtained since the results (both numerical and experimental) emanate from wave properties of the light.}
\end{figure}

By replacing the photosensitive surface by a convex lens (in experiment the lens was positioned before this point, however this has no great implication for the results) we can now obtain the distribution of the intensity/photons behind lens in the position where the geometrical optics predicts focused image of slits. The results are depicted on the Fig.~\ref{fig02}. When only one slit is open, it is clear that the incident photon originates from the open slit (upper slit thus corresponds to the lower peak and lower slit corresponds to the upper slit). When both slits are open, Afshar claims, that previous results still hold and he can thus obtain the which-way information. This is however false because having both slits open is a different experiment from the one with a single open slit. Numerical results confirm this, since to obtain the double-peaked interference pattern in Fig.~\ref{fig02} corresponding to the focused image of both slits, we used only wave properties of the light. The resulting image is thus an interference pattern that does not contain any which-way information.

\begin{figure}
\includegraphics[scale=0.32,angle=-90]{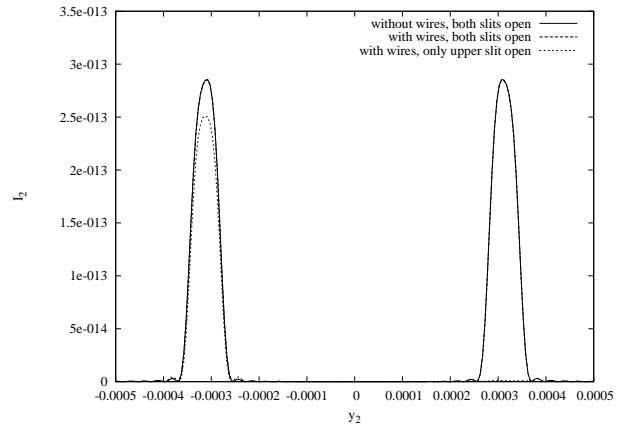}
\caption{\label{fig03}When the wires are placed on the interference pattern minima from the first simulation and both slits are open the resulting interferrence pattern bihind the lens stays the same. However when only one slit is open the wires cause the decrease in the light intensity since in this case there are no minima in intensity of light on the position of wires and this portion of light is intercepted.}
\end{figure}

To further support this claim we present results of simulation when the wires are put on the places where the interference pattern from the first simulation had its minima. Now it is clear that this cannot change results obtained without wires (see Fig.~\ref{fig03}) for both slits open, since the minima in the interference pattern mean that the probability of finding photon in thet place is zero and thus no photon is intercepted. Situation of course changes when only one slit stays open. Since there is a non zero probability to find a photon in the positiion of the wires, thes intercept some of the light so the obtained image has lower peak intensity than that obtained without wires. This decrease in intensity is as large as 10\% for collected data.

\section{Conclusion}
In conclusion, we have numerically simulated experiment from \cite{Afshar_old}. Since in the simulation only wave properties of light were used and the results correspond to those obtained by Afshar, we conclude that his claim, that he can obtain the which-way information from the position of the photon incident on the photosensitive surface placed behind the lens, is false. The which-way information in the case of both slits open is thus unobtainable and the same probably holds also for Afshar's second experiment carried out in \cite{Afshar_new}.


\vfill

\begin{thebibliography}{0}

\bibitem{Afshar_old} Shahriar S.~Afshar: \emph{Violation of the principle of Complementarity, and its implications}, Proc.~SPIE {\bf 5866} (2005) 229-244, {\tt quant-ph/0701027}

\bibitem{Afshar_new} Shahriar S.~Afshar: \emph{Violation of Bohr's Complementarity: One Slit or Both?}, AIP Cof.~Proc. {\bf 810}, (2006) 294-299, {\tt quant-ph/0701039}

\bibitem{Tabish} Tabish Qureshi: \emph{Complementarity and the Afshar Experiment}, {\tt quant-ph/0701109}

\end{thebibliography}
\end{document}